\documentclass[nofootinbib,notitlepage,superscriptaddress,10pt,aps,prd,twocolumn]{revtex4-1}

\usepackage{amsmath,mathtools}
\usepackage{tensor}
\usepackage{amssymb}
\usepackage{graphicx}
\usepackage{verbatim}

\newcommand{\leri}[1]{\left(#1\right)}

\DeclareMathOperator{\sgn}{sgn}
\begin{document}

\preprint{APS/123-QED}

\title{Semi-classical and quantum 
analysis of the isotropic Universe in the 
polymer paradigm}

\author{Giovanni Montani}
\email{giovanni.montani@enea.it}
\affiliation{ENEA, FSN-FUSPHY-TSM, R.C. Frascati, Via E. Fermi 45, 00044 Frascati, Italy}
\affiliation{Physics Department, ``Sapienza'' University of Rome, P.le Aldo Moro 5, 00185 (Roma), Italy}
\author{Claudia Mantero}
\email{claudia.mantero@uniroma1.it}
\affiliation{Physics Department, ``Sapienza'' University of Rome, P.le Aldo Moro 5, 00185 (Roma), Italy}
\author{Flavio Bombacigno}
\email{flavio.bombacigno@uniroma1.it}
\affiliation{Physics Department, ``Sapienza'' University of Rome, P.le Aldo Moro 5, 00185 (Roma), Italy}
\author{Francesco Cianfrani}
\email{Francesco.Cianfrani@uniroma2.it}
\affiliation{University of Rome Tor Vergata, Department of Industrial Engineering, Via del Politecnico 1, Rome 00133, Italy}
\author{Gabriele Barca}
\email{barca.1608557@studenti.uniroma1.it}
\affiliation{Physics Department, ``Sapienza'' University of Rome, P.le Aldo Moro 5, 00185 (Roma), Italy}

\date{\today}

\begin{abstract}
We analyse the semi-classical and quantum dynamics of the isotropic Universe in the framework of the Polymer Quantum mechanics, in order to implement a cut-off physics on the initial singularity. 
We first identify in the Universe cubed scale factor (\textit{i.e.} the spatial volume) the suitable configuration variable, providing a constant critical energy density, such that the Bounce arises as intrinsic geometric feature. We then investigate the obtained semi-classical Bounce dynamics for the primordial Universe, and we outline its impact on the resolution of cosmological paradoxes, as soon as the semi-classical evolution is extended (in the spirit of the Ehrenfest theorem) to the collapsing pre-Bounce Universe. Finally, we validate the use of the semi-classical effective dynamics by investigating the behaviour of the expectation values of a proper semiclassical states. The present analysis has the merit to enforce the equivalence between the Polymer quantization paradigm in the Minisuperspace and the Loop Quantum Cosmology approach. In fact, our study allows to define a precise correspondence between the Polymer cut-off scale and the discrete geometric structure of LQG. 
\end{abstract}

\pacs{Valid PACS appear here}
\maketitle

\section{Introduction}
The theory of Loop Quantum Gravity (LQG) \cite{Thiemann:2007zz} demonstrated that is possible to recover a notion of discrete space (discrete spectra for the kinematic areas and volumes operators \cite{Rovelli:1994ge}), even starting from a continuous Hamiltonian representation. In this regard, the main achievement of this formulation is the elimination of the singularity in the Friedmann-Lema\^{i}tre-Robertson-Walker (FLRW) Universe. In particular, as outlined in \cite{Ashtekar:2006uz,Ashtekar:2006rx,Ashtekar:2006wn}, the emergence of a minimum value for cosmic scale factor is responsible for the appearing of a Bounce scenario.
\\ \indent Despite a more general implementation of the symmetry 
prescriptions (homogeneity and isotropy) within the 
$SU(2)$ gauge structure of the space-time 
is to be defined, see \cite{Cianfrani:2011wg,Cianfrani:2010ji,Cianfrani:2012gv,Alesci:2015nja,Alesci:2016xqa}, the Big-Bounce paradigm opens 
a new perspective in Primordial Cosmology \cite{Primordial}.
\\ \indent Polymer quantum mechanics has been introduced 
in \cite{Ashtekar:1995mu,Ashtekar:2000eq}
by close analogy with the formulation of Loop Quantum 
Cosmology (LQC), since the former reproduces some key 
features of the latter, like the Hilbert space 
structure and the semi-classical dynamics. 

However, as discussed in \cite{Corichi:2006qf,G.:2014lpa} the Polymer formulation 
of the Minisuperspace dynamics is also an independent quantization procedure, 
able to describe cut-off physics effects in the 
cosmological setting. Therefore, in view of a comparison in the semi-classical limit with the Loop predictions, it can be instructive to wonder which are the most suitable configuration variables for quantization. In fact, standard canonical variable do not provide equivalent quantum systems when the Polymer prescription 
for the Polymer momentum operator is implemented. This is due to the violation of one of the hypotheses of the Stone-Von Neumann theorem \cite{SVN}, namely weak-continuity, which makes non unitarily equivalent those quantum descriptions based on different choices of the configuration variables.
 Thus, while LQC finds in the 
Ashtekar-Barbero-Immirzi variables its natural 
implementation, the Polymer prescription contains 
a non-trivial degree of ambiguity. 
\\ In particular, by requiring that the revised Friedmann 
equation retains the same form of the semi-classical effective LQC equation, with the energy density cut-off depending only on the Polymer 
parameter, the cubed scale factor is selected as the proper dynamical variable. Clearly, LQC effective equations can be found also for a generic choice of the Minisuperspace variable (in this sense it naturally contains the Polymer formulation), but at the price to deal with a Polymer parameter depending on the Universe scale factor. 
In this sense, the analysis below demonstrates that 
a change of variable in Polymer quantum cosmology 
corresponds to a redefinition of the discretization 
parameter as a given function of that variable.
Nonetheless, the idea of a constant cut-off scale 
seems to be a privileged choice because fixes the energy density cut-off as an intrinsic property of the theory and establishes a direct link with the Immirzi parameter of LQG.\\ 
\\ \indent We first consider the scale factor as the most natural variable, showing that it provides a representation of the Universe dynamics, characterized by a Bounce scenario only for supra-radiation equation of state, \textit{i.e.} $p > \rho /3$, $p$ and $\rho$ being the Universe pressure and energy density, respectively. Furthermore, we outline that, also for the case of a flat spatial geometry, a turning point in the future appears, strictly related to the value of the Polymer parameter associated to the lattice step.
\\ These two unpleasant features lead us to search for a suitable configuration variable, such that polymer quantization predicts a bounce, whose features are independent of the matter filling the space and it can be interpreted as an intrinsic geometrodynamical property of the considered quantum gravity approach. 
\\ \indent We identify such a variable in the cubed scale factor, which characterizes the geometrical volume of the Universe and therefore seems to have a privileged dynamical role. Then, we analyse in detail the features of the obtained bounce cosmology, demonstrating the validity of the usual 
conservation law for the energy density and the divergence of the cosmological horizon, as soon as the pre-Bounce evolution is taken into account.
\\ We also discuss the status of the horizon and flatness paradoxes in polymer cosmology with the adopted set of variables. We outline that the former is naturally solved, since during the bouncing era those regions which are now causally disconnected were in causal contact. Concerning the latter, we trace the evolution of the curvature parameter, which remains finite across the bounce and in the pre-bouncing phase it follows the classical trajectory for a contracting Universe. The issue of initial conditions is thus moved from the initial singularity to the pre-bounce classical phase, reducing the required amount of fine-tuning on the curvature parameter. 
Finally, we provide a pure quantum implementation of the considered scenario for an inflationary paradigm, \textit{i.e.} including in the Universe dynamics the energy density of a scalar field (the inflaton kinetic energy) and a positive cosmological constant (representing the false vacuum energy in the slow-rolling phase
 \cite{Primordial,Kolb:1990vq}). This analysis has the main purpose to validate, in the sense of the Ehrenfest theorem, the semi-classical equations used in the previous studies in a relevant cosmological context (they are legitimated by the predictions of Loop Quantum Cosmology (LQC) too \cite{Ashtekar:2007tv,Ashtekar:2011ni}).
\\ \indent The same model has been considered in a slightly different framework by \cite{Pawlowski:2011zf,Kaminski:2009pc}, where the problem of the non-self-adjointness of the Hamiltonian is addressed. Since we are interested in the Bounce dynamics, the conjugate momentum of the cubed scale factor is much greater than the cosmological constant value. Therefore, in the considered momentum representation 
(the only viable in the Polymer continuum limit), the Hamiltonian approaches an Hermitian operator.
\\ \indent The behavior of the quantum packet is compared with the analytical solution, obtained in the semi-classical limit for low momentum value with respect the cut-off: In this respect our comparison between quantum and semi-classical features is strictly valid for momentum values between the inverse of the lattice step and the squared inverse of the cosmological constant. In this region, as shown in \cite{Bojowald:2010qm}, the wave packet may deform, but effective semiclassical equations capture the main features of the quantum dynamics. Hence, the Ehrenfest theorem qualitatively holds, encouraging the idea that the semi-classical dynamics concerns also the passage of the Universe across the Bounce configuration. This opens a new point of view on the origin of the Universe thermal history and on the solution of its paradoxes.
\\ \indent Eventually, we want to stress that the choice of 
the cubed scale factor as dynamical variable, first reported in \cite{Ashtekar:2007em} as ``simplified LQC'', allows a direct comparison of the obtained modified Friedman equation with the one proper of LQC. In fact, the two equations retain exactly the same form, which permits to fix a precise link between the Immirzi parameter and the Polymer cut-off value. 
\\ The manuscript is organized as follows. In section \ref{II} polymer quantization of the scale factor is performed and the emergence of a matter-dependent critical energy density is emphasized. Then, in section \ref{III} the suitable change of variables is derived such that the critical energy density becomes independent of the dynamics and the resulting energy cut-off is related with the discretization scale and the Immirzi parameter. In section \ref{IV} the horizon and flatness paradoxes are critically discussed, while in section \ref{V} a deparametrized model for the inflaton is considered and the resulting dynamics for expectation values is numerically integrated. Finally, brief concluding remarks follow in section \ref{VI}.

\section{Polymer cosmology: discretization of the scale factor}\label{II}
Let us consider the homogeneous and isotropic universe described by the FRWL line element\footnote{In the following we set the speed of light $c=1$.}
\begin{equation}
ds^2=N(t)^2\,dt^2-a(t)^2\left(\dfrac{dr^2}{1-kr^2}+r^2d\theta^2+r^2sin\theta^2d\phi^2\right),\label{metricafrw}
\end{equation} 

where $k$ represents the curvature of the spatial hypersurface, $N(t)$  the lapse function and $a(t)$  the scale factor. We note that the lapse function $N(t)$ is a Lagrange multiplier related to the choice of the time variable, therefore  $a(t)$ can be considered the only actual dynamical degree of freedom.
\\ \indent In the presence of a density $\rho$ given by
\begin{equation}
\rho(a)=\dfrac{\mu^2}{a(t)^{3(w+1)}},\label{rhodia}
\end{equation}
$\mu$ being a constant and $w$ denoting the so-called polytropic index relating pressure and density, \textit{i.e.} $p=w\rho$, the super-Hamiltonian of the model reads as:

\begin{equation}
H=-\dfrac{ 2 \pi G p_a^2}{3\,a}-\dfrac{3    ka}{8 \pi G}+\dfrac{\mu^2}{a^{3w}},\label{vincolo11}
\end{equation}

where $p_a$ is the conjugate momentum of $a(t)$. 
\\ \indent In PQM a fundamental discrete structure for the configuration variable is implemented by the choice of the Hilbert space, namely $\mathcal{H}_{poly}$ $= L_2(\mathbb{R}_d,d\mu_d)$, where $d\mu_d$ is the Haar measure and $\mathbb{R}_d$ denotes the real line endowed with a discrete topology
 \cite{Ashtekar:2003hd}. In particular, the polymer framework is properly described by a dimension-full parameter\footnote{In particular, the parameter $\lambda$ has dimension of $[L]^3$} $\lambda$ such that the standard  Schr\"{o}dinger representation is recovered in the continuum limit $\lambda \rightarrow 0$ (see \cite{Corichi:2007tf,Corichi:2007hg}).\\ 
\indent Notably, if we assume that the scale factor $a(t)$ is discrete, with lattice length $\lambda$, it can be seen that the associated momentum operator $\widehat{p}_a$ does not exist. Therefore, with the aim of capturing the main semi-classical effects, it can be performed the replacement
\begin{equation}
p \rightarrow \frac{\hbar}{\lambda}\sin \left(\frac{\lambda p}{\hbar}\right),
\label{subst}
\end{equation}
that can be demonstrated \cite{Ashtekar:2006wn} to be a valid approximation in the continuum limit.
\\ \indent By means of \eqref{subst}, an effective Hamiltonian can be inferred from \eqref{vincolo11}, namely:
\begin{equation}
H_{poly}=-\dfrac{ {2 \pi \hbar^2 G}}{3  \lambda^2 a} \sin^2\left(\frac{\lambda p_a}{\hbar}\right)-\dfrac{3ka}{8\pi G}+\dfrac{\mu^2}{a^{3w}},\label{vincolo1}
\end{equation}
which represents the starting point of our analysis.
\\ \indent Hence, properly combining the equations of motion stemming from \eqref{vincolo1} in the synchronous reference, it is easy to obtain a modified Friedman equation for $k=0$, that is:
\begin{equation}
\left(\frac{{a'}(\tau)}{a(\tau)}\right)^2=\frac{8 \pi Q}{3}\left(1- \frac{Q}{Q_c}\right),\label{atauadim}
\end{equation}


where a prime denotes differentiation with respect the argument and we defined the dimensionless quantities\\
\begin{equation}
 \tau=\dfrac{t}{t_{Pl}},\quad Q=\dfrac{\rho}{\rho_{Pl}},\quad Q_c=\dfrac{\rho_c}{\rho_{Pl}},
\end{equation}
with $t_{Pl}$ and $\rho_{Pl}$  the Planck time and the Planck density, respectively.\\
In particular, we introduced the critical density $\rho_c$, defined in terms of the scale factor $a(\tau)$ and the polymer parameter $\lambda$, namely:
\begin{equation}
\rho_c=\dfrac{2\pi\hbar^2 G }{3 \lambda^2 a(\tau)^4}.
\label{rhocrit}
\end{equation}

\begin{figure}[h!]
\begin{center}
\includegraphics[width=\columnwidth]{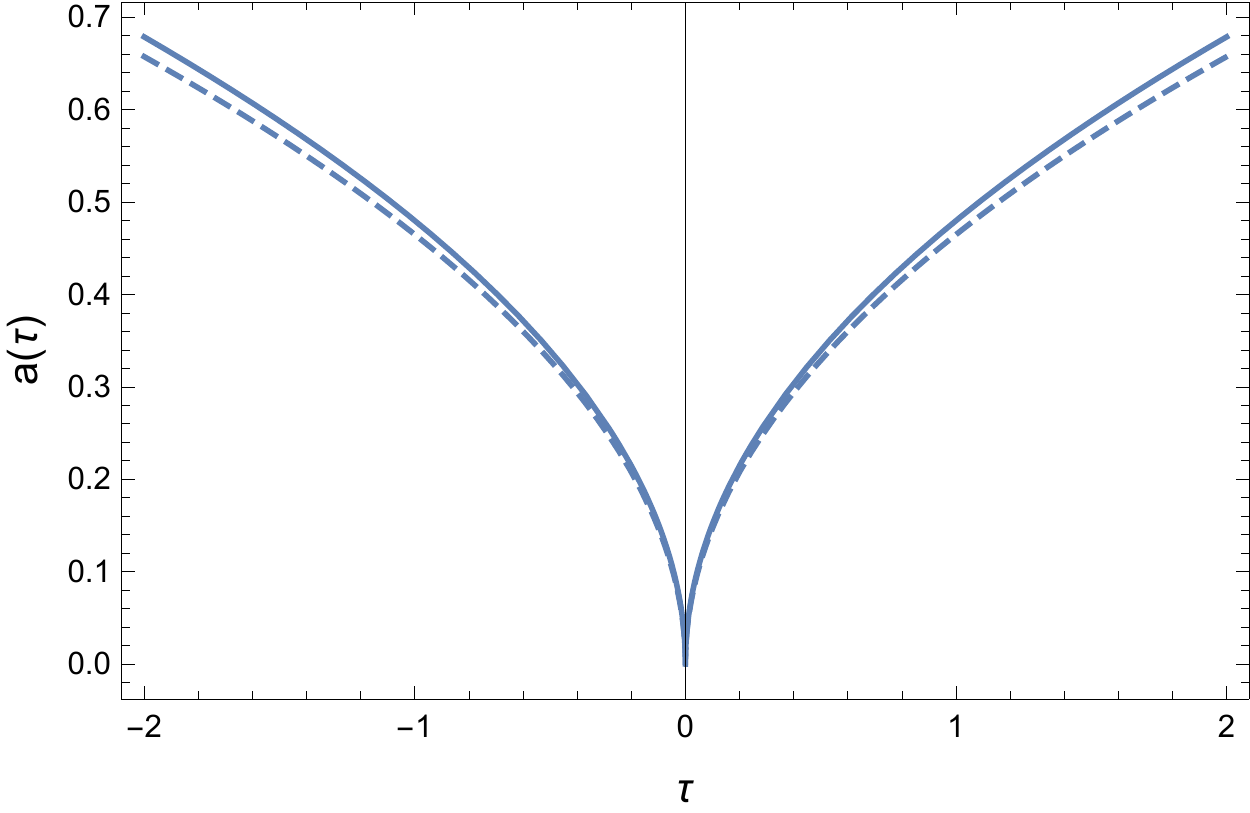}
\end{center}
\caption{\footnotesize Comparison between the standard (continuous line) trajectory and the polymer (dashed line) trajectory of $a(t)$  for $\lambda= 0.1$, $\mu=1$ and $w=\frac{1}{3}$ (with $\hbar=c=8 \pi G=1$). The initial singularity is still present and the scale factor shrinks to $0$ for $t=0$.}
\label{fig:a13k0a}
\end{figure}

\begin{figure}[h!]
\begin{center}
\includegraphics[width=\columnwidth]{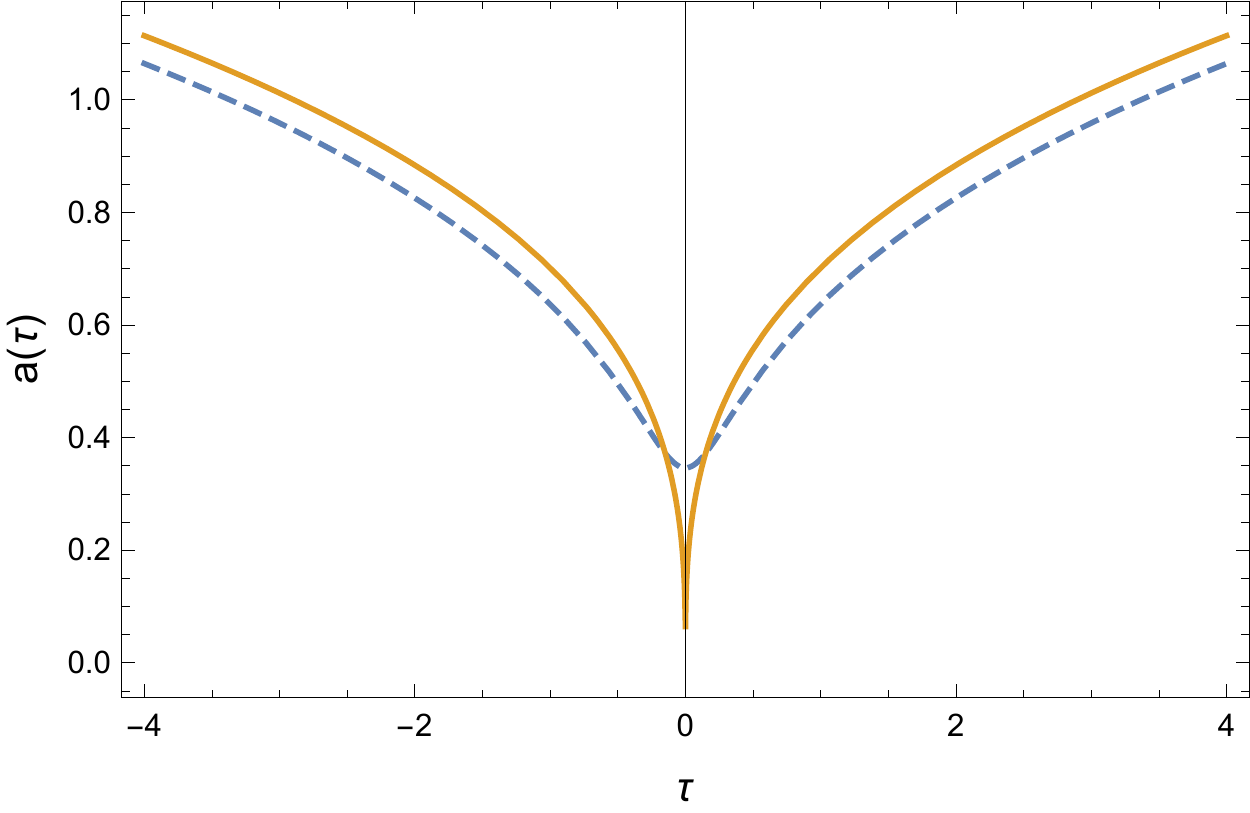}
\end{center}
\caption{\footnotesize Comparison between the standard (continuous line) trajectory and the polymer (dashed line) trajectory of $a(t)$  for $\lambda= 0.1$, $\mu=1$ and $w=1$ (with $\hbar=c=8 \pi G=1$). }
\label{fig:a1k0a}
\end{figure}

The main implication of the modified Friedmann equation \eqref{atauadim} is the existence of bounce and turning points for the scale factor evolution, occurring at $a=a^*$, with\footnote{Provided that $w \neq \frac{1}{3}$}
\begin{equation}
a_*=\left(\frac{3 \lambda ^2 \mu ^2}{2\pi \hbar^2 G}\right)^{\frac{1}{3 w-1}}.
\end{equation}
In this framework, no bounce is predicted for a radiation-dominated Universe, $w=\frac{1}{3}$ (Fig.\ref{fig:a13k0a}) and the classic and the polymer-modified trajectories overlap each other. In this respect, it is worth stressing that the analysis in \cite{Pawlowski:2014fba} shows the existence in LQC of a bouncing scenario also in the case of a radiation dominated Universe, in close analogy to what originally discussed for the case of a massless scalar field (stiff-matter case). Then again, for a stiff matter-dominated Universe, \textit{i.e.} $w=1$, a bouncing point appears (Fig.\ref{fig:a1k0a}) and the scale factor shows no initial singularity.

Therefore, by virtue of \eqref{rhocrit}, the position of the bounce is determined by the matter filling the Universe. Then, with the aim of obtaining bounce and turning points related to geometrical properties only, we look for a canonical transformation which allows us to infer a critical density independent from the scale factor. The same issue is solved in LQC by the so-called $\bar{\mu}$ regularization scheme \cite{Ashtekar:2006uz}, in which the holonomy correction is assumed to be metric-dependent in order to derive a constant critical energy density $\rho_c$. The analysis below achieves the same result from the point of view of polymer quantization, fixing a proper correspondence between PQM in Minisuperspace and the $\bar{\mu}$ regularization scheme in LQC. Alternatively, to restore an isomorphism with LQC we can consider a polymer parameter depending on the cosmic scale factor, {\it i.e.} $\lambda=\lambda(a)$.  This suggests that in PQM a change of variables can be interpreted as a dynamical discretization scale, even though PQM and LQC still remain independent quantization techniques. Indeed, LQC is grounded upon peculiar spin networks classes stemming from a classical reduction of the degrees of freedom of the full theory of LQG. Instead, the polymer approach represents an insightful tool for implementing at the
kinematic level cut-off physics effects.
 
\section{Optimized Polymer Cosmology and the link with LQC}\label{III}
Let us introduce a new configuration variable $A(\tau)$ defined by
\begin{equation}
A(\tau)\equiv f(a(\tau)),
\label{fdia}
\end{equation}
where $f(a)$ is a generic function of the scale factor and whose conjugate momentum reads as:
\begin{equation}
P_A=\frac{p_a}{f'(a)}.
\label{canonicaltr}
\end{equation}

\begin{figure}[h!]
\begin{center}
\includegraphics[width=\columnwidth]{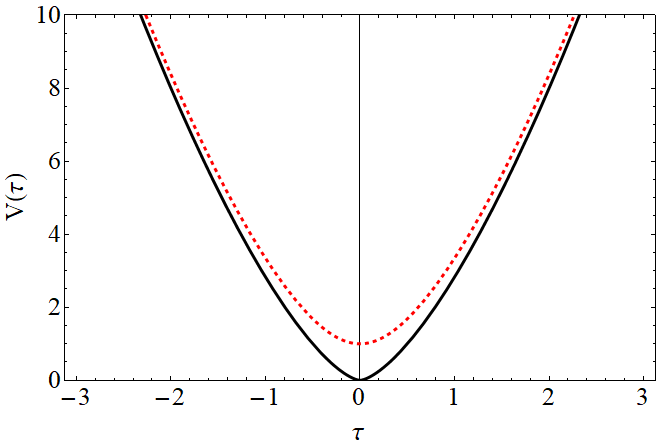}
\end{center}
\caption{\footnotesize Comparison between the standard (continuous line) trajectory and the polymer (dashed line) trajectory of $V(\tau)$.}
\label{fig:V_flat_rad}
\end{figure}

\begin{figure}[h!]
\begin{center}
\includegraphics[width=\columnwidth]{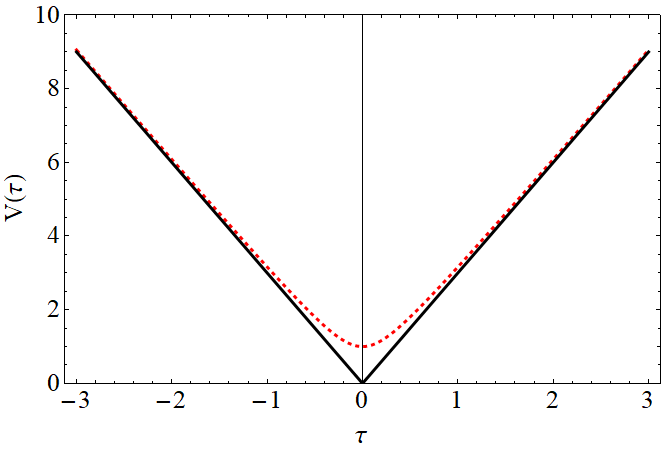}
\end{center}
\caption{\footnotesize Comparison between the standard (continuous line) trajectory and the polymer (dashed line) trajectory of $V(\tau)$. }
\label{fig:V_flat_stiff}
\end{figure}

Thus, inserting \eqref{canonicaltr} in the standard Hamiltonian \eqref{vincolo11} and performing the polymer approximation \eqref{subst} yields
\begin{equation}
H_{poly}=-\frac{2 \pi \hbar^2 G \left(f'(a)\right)^2}{3   \lambda^2 a}\sin^2\left(\frac{\lambda P_A}{\hbar}\right)-\dfrac{3  k a}{8 \pi G}+\dfrac{ \mu^2}{a^{3w}}.
\end{equation}\\
The analogous of \eqref{atauadim} becomes, 
\begin{equation}
\left(\frac{{a'}(\tau)}{a(\tau)}\right)^2=\frac{8 \pi Q}{3}\left(1- \frac{Q}{\tilde{Q}_c}\right),\label{frimod}
\end{equation}

where we defined
\begin{equation}
\tilde{Q}_c\equiv f'(a)^2 Q_c.
\end{equation}
Now, if we require that the critical density is independent of the scale factor, we get the condition
\begin{equation}
\frac{ a^4}{  f'(a)^2}=C,
\end{equation}
with C a constant, which admits the solution
\begin{equation}
f(a)\sim a^3\equiv V.
\end{equation}
Since the Einstein equations are modified by the polymer assumption \eqref{subst}, we expect that also the relation $\rho\equiv \rho(V)$, that is the continuity equation (which we implicitly used in \eqref{rhodia}) could be altered due corrections of $\lambda$ order.
However, solving the Hamilton constraint with respect to $P$ and manipulating the Hamilton and Friedman equations, the continuity relation can be rearranged in the following way:
\begin{equation}
\rho'(t)=-\dfrac{V'(t)}{V(t)}\left(\rho-\dfrac{d(\rho V)}{dV}\right),\label{continuity equation}
\end{equation}
which is identically satisfied by the standard expression
\begin{equation}
\rho(V)=\dfrac{\mu^2}{V^{w+1}},\label{density on V}
\end{equation}
that rules out any corrections of $\lambda$-order for the energy density $\rho$.
\\ \indent Hence, using $\{V,P\equiv P_A\}$ as canonical variables, a Friedman-like equation for the polymer case can be inferred, that is 
\begin{equation}
\left(\dfrac{{V'}(\tau)}{{V}(\tau)}\right)^2=24\pi Q\left(1-\dfrac{Q}{\tilde{Q}_c}\right),\label{Vadim}
\end{equation}

where  $\tilde{Q}_c=\dfrac{\tilde{\rho}_c}{\rho_{Pl}}$ and 
\begin{equation}
\tilde{\rho}_c=\dfrac{6 \pi \hbar^2 G}{  \lambda^2},\label{rhocritpoly}
\end{equation}
and from \eqref{Vadim} it can be seen as the critical points are now given by
\begin{equation}
V_*=\left(\frac{  \lambda^2\mu^2}{6 \pi  \hbar^2G}\right)^{\frac{1}{w+1}}.
\label{bounceV}
\end{equation}
Now, for the radiation-dominated Universe the critical points \eqref{bounceV} do not exhibit any ambiguities in $w=\frac{1}{3}$, and a bouncing solution can be obtained (Fig.\ref{fig:V_flat_rad}). Lastly, regarding the stiff matter-dominated Universe, the initial singularity is still avoided (Fig.\ref{fig:V_flat_stiff}).

\indent It is remarkable that \eqref{Vadim} is in agreement with the result obtained in LQC \cite{Ashtekar:2006wn} (see also the seminal paper \cite{Bojowald:2006gr}, in which the robustness of this scenario is outlined in the presence of a massless scalar field), where the critical density is actually given by
\begin{equation}
\rho_c=\dfrac{\sqrt{3}}{16\pi^2 \hbar G^2 \gamma^3},\label{rhocritloop}
\end{equation}
with $\gamma$ the Immirzi parameter \cite{Immirzi:1996di,Immirzi:1996dr,Rovelli:1997na}. Nonetheless, we underline that for a flat FRLW model the Immirzi parameter disappears from the classical phase space and it enters the quantum dynamics only with the ``ad-hoc'' assumption that the minimal area gap of LQC must equal that of full LQG (see for instance \cite{BenAchour:2017qpb,BenAchour:2018jwq}). In particular, in \cite{BenAchour:2017qpb} it is shown how the classical symmetry changing the value of the Immirzi parameter can be implemented on a quantum level by adopting the Thiemann complexifier technique. Furthermore, in \cite{BenAchour:2018jwq} the LQC problem is addressed by reconciling a minimal length scale with fundamental space-time symmetries. Our comparison between the Immirzi parameter and the polymer discretization scale follows the original idea that the minimal area gap in LQC and LQG must coincide. However, the following analysis is independent of such relation and it characterizes the Universe behavior in a consistent polymer approach, able to ensure a bounce feature independently of the matter equation of state. In particular, comparing \eqref{rhocritpoly} and \eqref{rhocritloop}, the parameter $\lambda$ can be related to $\gamma$, namely:
\begin{equation}
\lambda=4\sqrt{2} \pi \l_P^3\gamma^{3/2} ,
\end{equation}
being $l_P$ the Planck length. \\
Such a result points out the consistency of picking $V(\tau)$ as the suitable dynamical variable and it sheds new light on a likely connection between the semi-classical descriptions provided by both polymer and LQC theories, which deserves further investigation. \\
\indent In summary, we claim that the use of the canonical variables $ (P,V )$ provides the proper theoretical setting for analysing the dynamical properties of the Universe within the polymer approach.  Indeed, it allows us to extend the range of applicability of the model to the radiation case and to solve the issues related to the definition of the critical density.\\

\section{Analysis of the horizon and flatness paradoxes}\label{IV}
\begin{figure}[h!]
\begin{center}
\includegraphics[width=1\columnwidth]{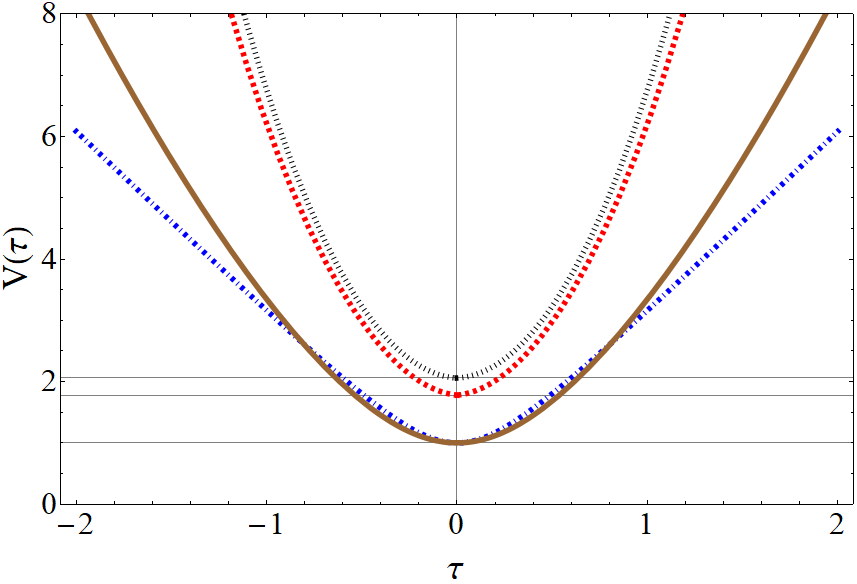}
\end{center}
\caption{\footnotesize Comparison between the polymer solutions for $\bar{k}=0$ and $\bar{k}<0$, in the presence of radiation ($w=1/3$) and stiff matter ($w=1$). In particular, the different cases are depicted by: solid line ($\bar{k}=0,\;w=1/3$), dot-dashed line ($\bar{k}=0,\;w=1$), dotted line ($\bar{k}<0,\;w=1/3$), dashed line ($\bar{k}<0,\;w=1$).}
\label{fig:V_PolyKneg}
\end{figure}

\begin{figure}[h!]
\begin{center}
\includegraphics[width=1\columnwidth]{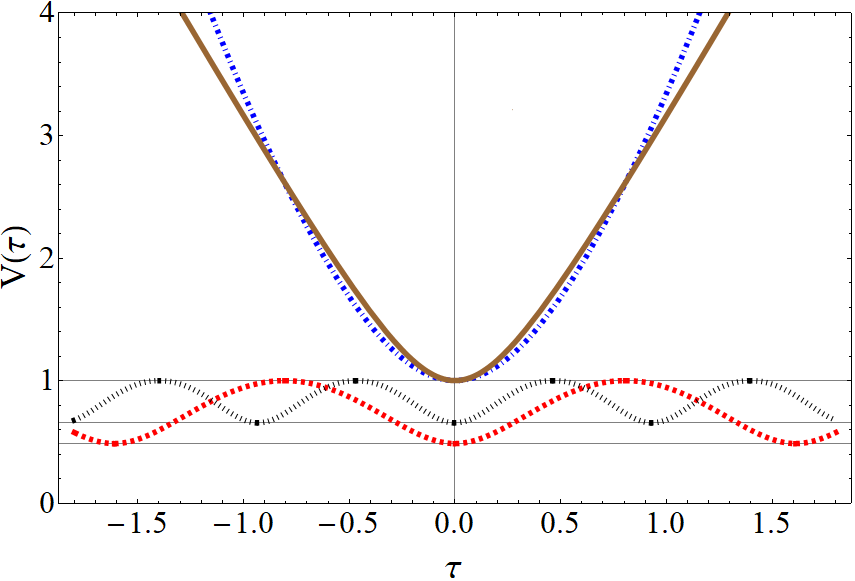}
\end{center}
\caption{\footnotesize Comparison between the polymer solutions for $\bar{k}=0$ and $\bar{k}>0$, in the presence of radiation ($w=1/3$) and stiff matter ($w=1$). In particular, the different cases are depicted by: solid line ($\bar{k}=0,\;w=1/3$), dot-dashed line ($\bar{k}=0,\;w=1$), dotted line ($\bar{k}>0,\;w=1/3$), dashed line ($\bar{k}>0,\;w=1$). The cyclical behaviour for positive curvature is highlighted by the maximal and minimal volumes.}
\label{fig:V_PolyKpos}
\end{figure}

It can be instructive to enlarge our analysis to include the case $k\neq 0$ as well, with the aim of study in detail the effect of the spatial curvature on the Universe evolution. Therefore, let us rewrite for the sake of convenience the Hamiltonian constraint \eqref{vincolo11} in terms of the new couple of canonical variables $\{P,V\}$, \textit{i.e.} 
\begin{equation}
\mathcal{H}=-6 \pi G V P^2-\frac{3}{8\pi G}kV^{1/3}+\rho V.
\label{Hamiltonian curvature}
\end{equation}
Then, taking into account \eqref{subst} for the conjugate momentum $P$, with a bit of algebra from \eqref{Hamiltonian curvature} a modified Friedman equation can be easily obtained, namely
\begin{equation}
H^2=\leri{\frac{8\pi G}{3}\rho-\frac{k}{V^{2/3}}}\leri{1-\frac{3}{8\pi G \tilde{\rho}_c}\leri{\frac{8\pi G}{3}\rho-\frac{k}{V^{2/3}}}},
\label{Friedmann equation polymer k}
\end{equation}
where the critical density $\tilde{\rho}_c$ is defined like \eqref{rhocritpoly} and $H\equiv\dot{V}/3V$. In this respect, it is worth noting that the bounce now corresponds to
\begin{equation}
\rho=\tilde{\rho}_c+\frac{3}{8\pi G}\frac{k}{V^{2/3}}.
\label{critical density k}
\end{equation} 
Furthermore, by the inspection of \eqref{Friedmann equation polymer k} is also clear that for positive curvature a turning point is predicted for
\begin{equation}
\rho=\frac{3}{8\pi G}\frac{k}{V^{2/3}},
\label{turning density k}
\end{equation}
where we expect the Universe to reach its maximal extension, before re-contracting. Then, since the relation \eqref{continuity equation} is still valid, the dependence of $\rho$ on the variable $V$ is given again by \eqref{density on V}. Hence, once fixed $w$, the relations \eqref{critical density k} and \eqref{turning density k} determine the critical points $V^*$ similarly to \eqref{bounceV}, even though an analytical solution is not always attainable for \eqref{critical density k}.
\\  \indent Now, by close analogy with \eqref{Vadim}, it is useful to recast \eqref{Friedmann equation polymer k} in the form
\begin{equation}
\bar{H}^2=\leri{\frac{8\pi Q}{3}-\frac{\bar{k}}{V^{2/3}}}\leri{1-\frac{3}{8\pi\tilde{Q}_c}\leri{\frac{8\pi Q}{3}-\frac{\bar{k}}{V^{2/3}}}},
\label{Friedmann equation polymer k adim}
\end{equation}
where we introduced the dimensionless spatial curvature $\bar{k}\equiv l_{Pl}^2\,k$ and Friedman parameter $\bar{H}=t_{Pl}^2 H$, respectively. 
\\Thus, by numerical integration of  \eqref{Friedmann equation polymer k adim} we can see that for $\bar{k}<0$ the evolution of $V(\tau)$ is quite similar to the flat case (Fig.~\ref{fig:V_PolyKneg}), except for the minimal value of the ``volume" where the bounce occurs. In fact, as it can be seen from \eqref{critical density k}, in the presence of  negative curvature the density at the Big bounce is lower than the flat case and the minimal volume is bigger. 
\\Instead, for $\bar{k}>0$, we deal with a cyclic dynamics characterized by the alternation of expanding and contracting phases (Fig.~\ref{fig:V_PolyKpos}), bounded by the density values \eqref{critical density k}, \eqref{turning density k}.

\begin{figure}[h!]
\begin{center}
\includegraphics[width=1\columnwidth]{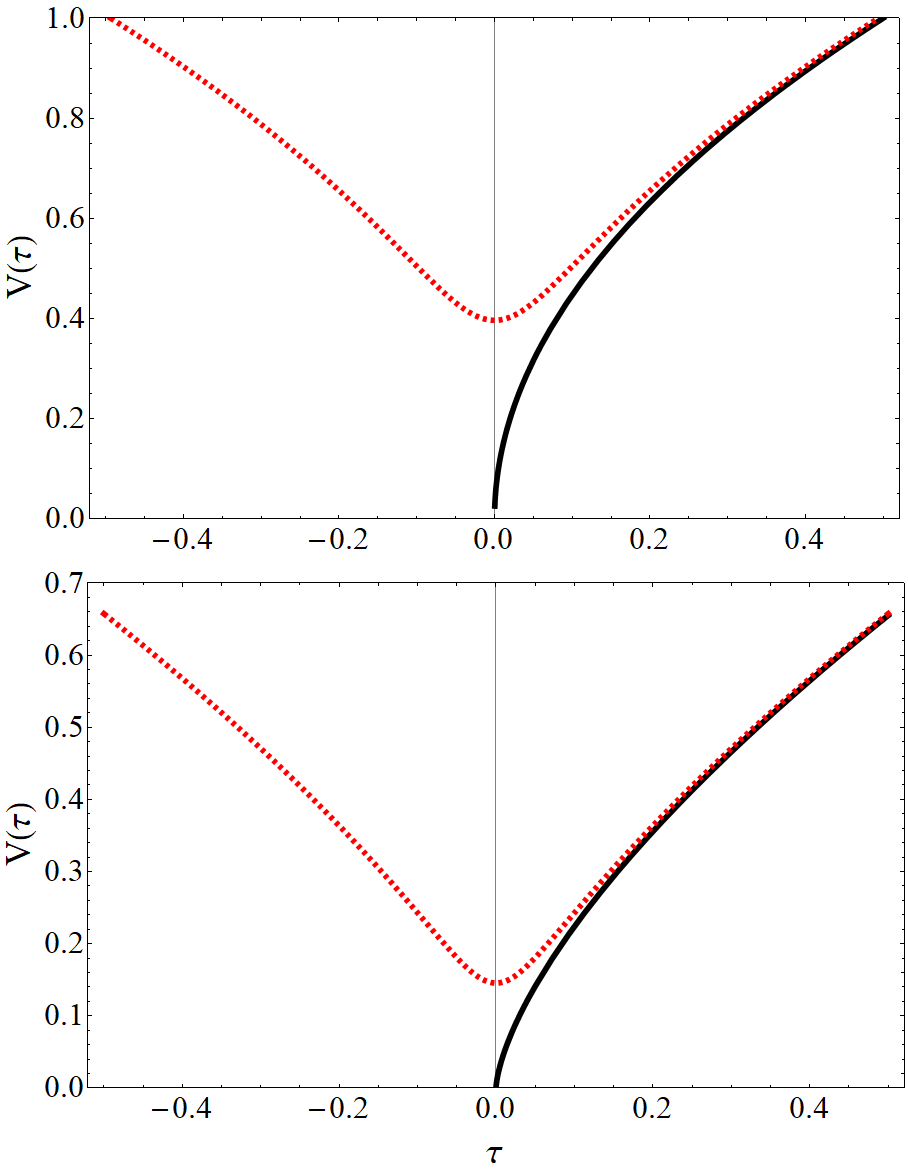}
\end{center}
\caption{\footnotesize Comparison between the classical (solid line) and the polymer (dashed line) behaviour of $\frac{d_h(\tau)}{V(\tau)^{1/3}}$, respectively, for $k=0$ in the presence of radiation $w=1/3$ (top) and stiff matter $w=1$ (bottom).}
\label{fig:ComovHorizonFlat_Tot}
\end{figure}

In the presence of the initial singularity the horizon paradox is a well-known shortcoming of the Standard Cosmological Model (SCM) \cite{Weinberg:2008zzc,Kolb:1990vq}. 
This paradox can be solved for a bouncing cosmology by the fact that disconnected causal regions in the expanding phase were causally connected during the pre-bounce contracting phase. Certainly, we note as such idea relies on the assumption that the semi-classical solution can be extended across the bounce, where quantum geometry effects were not negligible. 
\\ \indent Thus, in order to understand if region causally connected today had been in the same particle horizon in the past, we have to analyse the particle horizon in comparison with the scale of physical length ($\sim V^{1/3}$), namely:
\begin{equation}
\dfrac{d_h(\tau)}{V(\tau)^{1/3}}=\int_{\tau_0}^{\tau}\dfrac{ d\tau'}{V(\tau')^{1/3}},\label{horiz}
\end{equation}
where $\tau_0$ is a fiducial time, that in the classical case can be identified with the initial singularity.\\ \indent 
Classically, when $\tau \rightarrow \tau_0=0^+$, the quantity \eqref{horiz} vanishes identically. Instead, when a bounce appears in $\tau_0=0$ (Fig.\ref{fig:ComovHorizonFlat_Tot}), we are able to extend the integration boundary from any negative times up to a positive $\tau=T$ after the bounce. Therefore, in the limit of $\tau\rightarrow -\infty$ the quantity \eqref{horiz} diverges and we can say that those regions that now are causally disconnected, were actually in causal contact during the contracting phase of the Universe.

\begin{figure}[h!]
\begin{center}
\includegraphics[width=1\columnwidth]{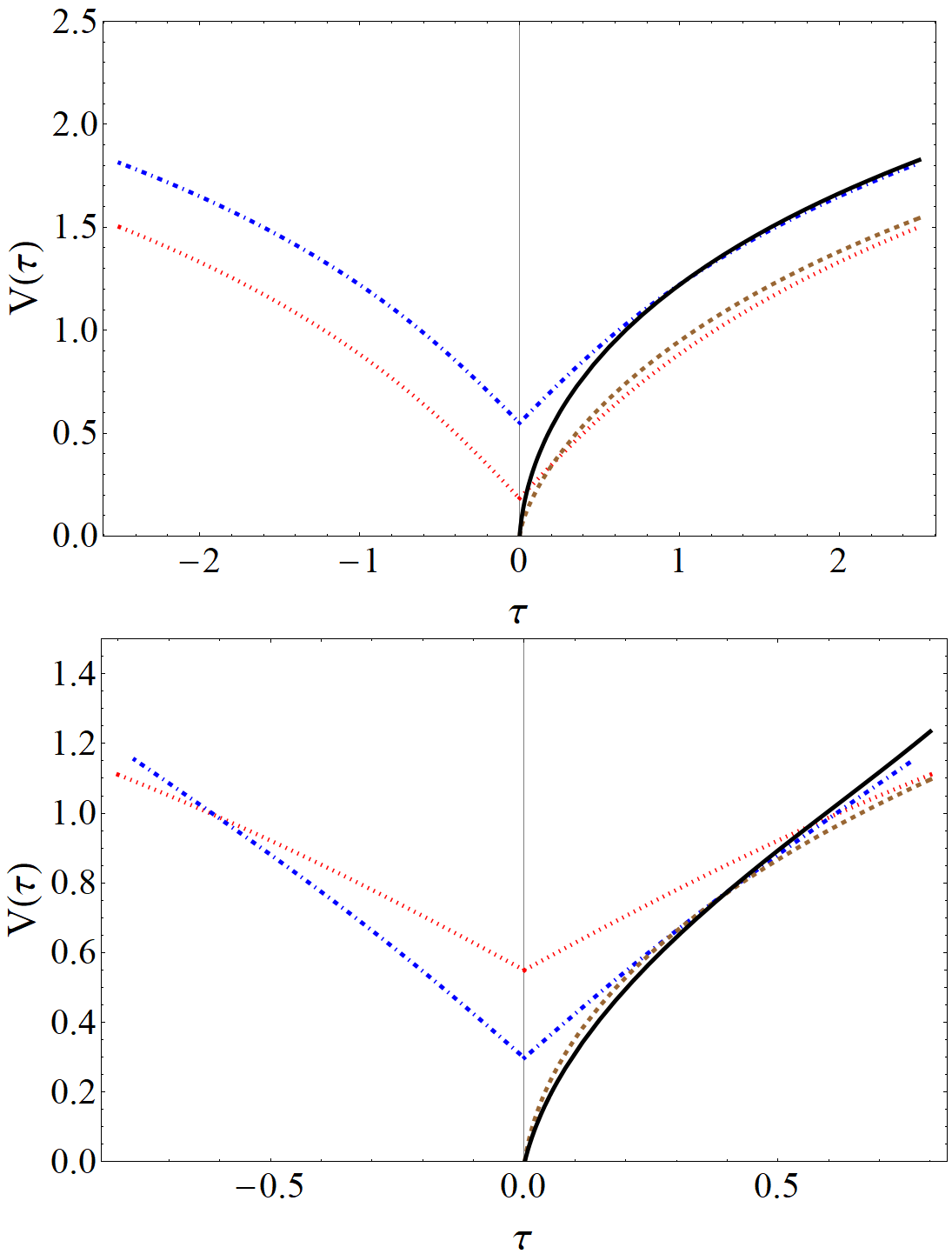}
\end{center}
\caption{\footnotesize Comparison between the classical and the polymer behaviours of $\frac{d_h(\tau)}{V(\tau)^{1/3}}$ for $\bar{k}<0 $ (top) and $\bar{k}>0$ (bottom). In particular, the different cases are depicted by: solid line ($w=1/3$, classic), dashed line ($w=1$, classic), dot-dashed line ($w=1/3$, polymer), dotted line ($w=1$, polymer). For the sake of clarity $\bar{k}$ is chosen much greater than its actual size, in order to outline properly the effects of the curvature.}
\label{fig:ComovHorizonK_Tot}
\end{figure}
\begin{figure}[h!]
\begin{center}
\includegraphics[width=1\columnwidth]{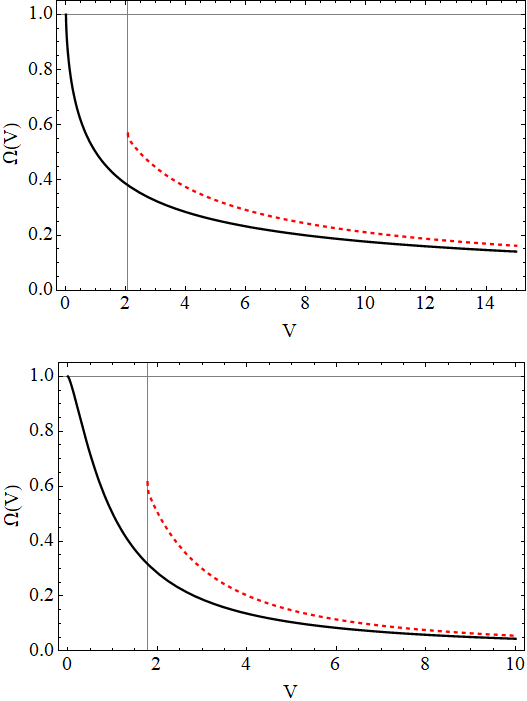}
\end{center}
\caption{\footnotesize The classical (solid line) and polymer (dashed line) behaviour of $\Omega$ for $\bar{k}<0$, when $w=1/3$ (top) and $w=1$ (bottom) cases are considered. For the sake of clarity $\bar{k}$ is chosen much greater than its actual size, in order to outline properly the effects of the curvature. The minimal value for the volume is also shown.}
\label{fig:Omega_Neg_Tot}
\end{figure}
\begin{figure}[h!]
\begin{center}
\includegraphics[width=1\columnwidth]{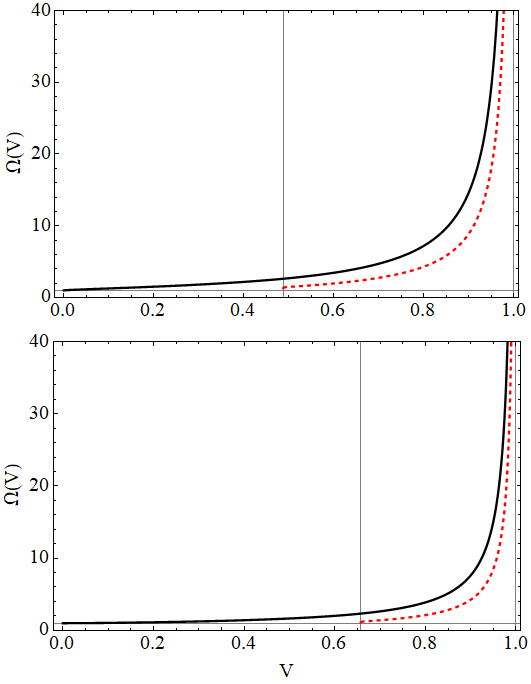}
\end{center}
\caption{\footnotesize The classical (solid line) and polymer (dashed line) behaviour of $\Omega$ for $\bar{k}>0$, when $w=1/3$ (top) and $w=1$ (bottom) cases are considered. For the sake of clarity $\bar{k}$ is chosen much greater than its actual size, in order to outline properly the effects of the curvature. The minimal value for the volume is also shown.}
\label{fig:Omega_Pos_Tot}
\end{figure}

It is worth remarking that such considerations are unaffected by the presence of the curvature, since also for a curved space (Fig.\ref{fig:ComovHorizonK_Tot}) the quantity \eqref{horiz} can be evaluated for $\tau_0\to -\infty$.
Thus, we see that in the present dynamical 
framework, we solve naturally the horizon paradox \cite{Primordial,Kolb:1990vq}. Indeed, if the considered dynamics remains valid across the Bounce (in the sense of the Ehrenfest theorem), the pre-existing collapsing Universe plays a role in fixing the spectrum of perturbations in the expanding one, and the inflationary paradigm could be revised (see also \cite{Brandenberger:2012um,Alexander:2015pda}).
\\ The issue of fine-tuning the initial conditions of the Universe, is mitigated but not completely solved in the present scenario. In order to see that, let us rearrange \eqref{Friedmann equation polymer k adim} into the form
\begin{equation}
\bar{H}^2=\frac{8\pi}{3}Q_k\leri{1-\frac{Q_k}{\tilde{Q}_c}},
\label{Friedmann equation polymer k adim flat-like}
\end{equation}
where we introduced the quantity
\begin{equation}
Q_k\equiv\frac{3}{8\pi}\leri{\frac{8\pi}{3}Q-\frac{\bar{k}}{V^{2/3}}},
\label{density k}
\end{equation}
that for $\bar{k}=0$ simply reduces to the ordinary density $Q$. Now, since for a classical Friedman Universe the density parameter $\Omega$ is simply given by
\begin{equation}
\Omega=\frac{8\pi Q}{3\bar{H}^2},
\label{density parameter class}
\end{equation}
it is natural, by virtue of \eqref{Friedmann equation polymer k adim flat-like}, to adopt for the polymer case the straightforward generalization of \eqref{density parameter class}, that is
\begin{equation}
\Omega=\frac{8\pi}{3\bar{H}^2}\;Q_{k=0}\leri{1-\frac{Q_{k=0}}{\tilde{Q}_c}}.
\label{density parameter poly 1}
\end{equation}
However, it can be verified that such a definition is actually misguided. In particular, it is easy to see that for the case $\bar{k}>0$, the relation \eqref{density parameter poly 1} leads to the conflicting condition $\Omega<0$. Indeed, if we denote with $V_{MIN}$ the volume at the bounce, in the presence of positive curvature the maximal density is given by (see \eqref{critical density k}):
\begin{equation}
Q_{MAX}=\tilde{Q}_c+\frac{3\bar{k}}{8\pi V_{MIN}^{2/3}}>\tilde{Q}_c,
\label{maximal density}
\end{equation}
and there exists a range of values for $Q$ where \eqref{density parameter poly 1} turns out to be negative. Thus, it is reasonable to modify \eqref{density parameter poly 1} into
\begin{equation}
\Omega=\frac{8\pi}{3\bar{H}^2}\;Q_{k=0}\leri{1-\frac{Q_{k=0}}{\tilde{Q}_{MAX}}}.
\label{density parameter poly 2}
\end{equation}
The numerical integration of \eqref{density parameter poly 2} shows that, for small values of $V(\tau)$, the behavior of the density parameter  in PQM slightly departs from the classical prediction (Fig.~\ref{fig:Omega_Neg_Tot}-\ref{fig:Omega_Pos_Tot}), and in correspondence of $V_{MIN}$ the condition $\Omega\sim 1$ still holds. Therefore, when the bounce is properly taken into account, the problem of fine-tuning the initial conditions is replaced by the demand for explaining why $\Omega$ is very close to $1$ today, even in the bouncing scenario. However, we stress that in \cite{Brandenberger:2016vhg} it is suggested that the flatness paradox could be understood in a bouncing cosmology as a dynamical effect due to the modified evolution of the scale factor, even if it undoubtedly deserves further investigations.

\section{Effective cosmological dynamics for physical states}\label{V}
\begin{figure}[h!]
\begin{center}
\includegraphics[width=\columnwidth]{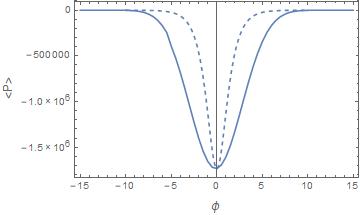}
\end{center}
\caption{\footnotesize Comparison between P($\phi$) (dashed line) and $<P(\phi)>$ (continuous line), numerically integrated for $\lambda= 10^{-6}$, $\Lambda=0.001$, $\sigma=0.22$, $\omega_*=200$ and $\phi_0=12$ .}
\label{fig:pipolymera}
\end{figure}

The Hamiltonian for General Relativity is weakly vanishing \cite{CanonicalQG}, and on a quantum level this implies the impossibility of defining an evolutionary operator, and quantum states do not seem to evolve.\\
On the other hand, it has been demonstrated that, by means of matter fields \cite{Cianfrani:2009vm,Giesel:2012rb,Blyth:1975is},  a notion of relational time can be restored and Schr\"{o}dinger-like equation written down.\\
\indent Thus, in order to mimic the inflationary paradigm, let us introduce into the model the scalar field $\phi$ and the cosmological constant $\Lambda>0$.
By means of $V$ the Hamiltonian in \eqref{vincolo11} can be rewritten as ($\hbar=c=8\pi G$=1):
\begin{equation}
H=-\frac{3}{4 }V P^2+{ V \Lambda}+ \frac{p_{\phi}^2}{4  V}=0,\label{hamilton}
\end{equation}
$p_{\phi}$ being the conjugate momentum of $\phi$. In order to use $\phi$ as internal time, we fix the time gauge
\begin{equation}
\dot{\phi}=1\implies N=\frac{2  V}{p_{\phi}},
\end{equation}
and solving \eqref{hamilton} with respect to $p_{\phi}$ we get the reduced Hamiltonian
\begin{equation}
p_{\phi}=-h_{rid}=\pm V {\sqrt{3{P}^2-4\Lambda}}\label{hrid}.
\end{equation}
Setting $p_{\phi}>0$, the solutions in the relational time $\phi$ for the equations of motion derived from \eqref{hrid} read as:
\begin{equation}
\begin{split}
P(\phi)&=\sqrt{\frac{4\Lambda}{3}}\sgn(\phi-\phi_0) \cosh (\sqrt{3}(\phi-\phi_0)),\\
V(\phi)&=\sqrt{\frac{3C}{4\Lambda}}\frac{1}{\vert \sinh(\sqrt{3}(\phi-\phi_0)) \vert},\label{PV}
\end{split}
\end{equation}
with $\phi_0$ and $C$ integration constants.\\
\indent Within the polymer scheme \eqref{subst}, if we perform an expansion up to the second order in $\lambda p$, the analogous of \eqref{hrid} is given by
\begin{equation}
p_{\phi}=-h_{rid}^{poly}=\pm V {\sqrt{3{P}^2-\lambda^2 P^4-4\Lambda}}.\label{hridpo}
\end{equation}
Although the equations of motion derived by \eqref{hridpo} do not admit analytical solutions, the junction point of the expanding and contracting branches for $\{P,\,V\}$ can be inferred, namely
\begin{equation}
V_b=\frac{\lambda  p_{\phi}}{\sqrt{3- 4 \lambda ^2 \Lambda }}\quad\quad P_b=-\sqrt{\frac{\sqrt{9-16 \lambda ^2 \Lambda }+3}{2\lambda ^2}},
\end{equation}
and in the limit $\lambda\rightarrow 0$ one can see that $V_b$ vanishes and the continuum picture is restored.
\\ \indent Our 
analysis concerns the region 
of the momentum space where the non-self-adjoint character of the Hamiltonian is mitigated, but some effects of the cosmological constant are still present. Therefore, with the aim of examining the evolution of the Universe at the Plank era, we study whether Dirac observables preserve a semi-classical evolution, \textit{i.e.} if the Ehrenfest theorem is violated. In particular, we have to compare the evolution of $P$ as predicted by \eqref{hridpo} with the expected value of the corresponding operator $\widehat{P}$.\\
Hence, assuming for $p_{\phi}$ and $V$ the standard representation the following Schr\"{o}dinger equation is obtained
\begin{equation}
\frac{\partial}{\partial \phi} \psi(P,\phi)=\left( \sqrt{3{ P}^2-\lambda^2 P^4-4\Lambda}\right)\frac{\partial}{\partial P}  \ \psi(P,\phi),  \label{shro}
\end{equation}
where $\psi$ is the wave function of the Universe, that can be put in the form
\begin{equation}
\psi(P,\phi)= \chi (P)\ e^{-i \omega \phi}.\label{sol}
\end{equation}
A solution for \eqref{sol} is given by
\begin{widetext}
\begin{equation}
\chi(P)=c_1 \exp \left(-\frac{\omega\;  \sqrt{\frac{\left(\lambda ^2-\frac{3}{2}\right) P^2}{4 \Lambda }+1}}{\sqrt{\frac{\lambda ^2}{f(\lambda )}} \sqrt{8 \Lambda+2 \lambda ^2 P^4-6 P^2}}\;F\left(i \sinh ^{-1}\left( i P \lambda\sqrt{\frac{2}{f(\lambda )}}\right)\vert \frac{3 f(\lambda )}{8 \lambda ^2 \Lambda }-1\right)\right),
\end{equation}
\end{widetext}
$c_1$ being an integration constant, $F(\cdot)$ denotes the hypergeometric function and $f(\lambda)$ is defined as
\begin{equation}
f(\lambda)=\sqrt{9-16 \lambda^2 \Lambda}+3.
\end{equation}
Semi-classical states can be constructed as wave packets associated to the wave function \eqref{sol}, peaked around classical values  $p_{\phi}= \omega_* > 0$ at a fixed time $\phi=\phi_0$, $\textit{i.e.}$
 \begin{equation}
 \Psi (P,\phi)= \int_{0}^{\infty} \frac{d\omega}{\sqrt{2 \omega}}\;\chi(P)\ e^{\frac{-(\omega-\omega_*)^2}{2 \sigma^2}}\ e^{-i \omega (\phi-\phi_0)}.
 \end{equation}
As shown in Fig.\ref{fig:pipolymera}, the comparison with the effective semi-classical dynamics for $\omega_*$ much smaller than the cut-off shows a qualitative good agreement: 
the mean value trajectory given by $<P>$ is well approximated by the Polymer modified classical one. The good agreement nearby the bounce region could be enhanced by a numerical analysis of the exact semi-classical equation, allowing for arbitrary momentum values.
\\ \indent Nonetheless, based on the results in \cite{Bojowald:2010qm}, in which higher moments are included, we can firmly claim that the Ehrenfest theorem is reliably valid near the Bounce. Therefore, we can extend the effective semiclassical formulation across the bounce and investigate the implications of the pre-bounce dynamics on the subsequent phases of the Universe, even though we have to justify the use of peaked wave packets in such extreme regime. We stress that the analysis in \cite{Bojowald:2010qm} is coherent with the present one, because the both consider minisuperspace models in the presence of a cosmological constant. In particular, it numerically demonstrates that, even if the form of a Gaussian packet can be significantly altered, the probability distribution singles out a precise hierarchy in higher order moments, allowing a truncation procedure. Thus, the study of the (finite) mean square root of the considered trajectories and the analysis in \cite{Bojowald:2010qm} permit us to justify the use of semi-classical equations for the predictions of the averaged quantum dynamics.

\section{Concluding remarks}\label{VI}
The possibility for a bouncing cosmology, both in a canonical scheme \cite{Cianfrani:2014iwa,Moriconi:2016egx}, and in a Polymer scenario \cite{Montani:2018bxv,Moriconi:2017bvs,Lecian:2013rea,Corichi:2007am}, provides a new cosmological point of view on the Universe birth. Whereas the singularity is a peculiar feature of the Einstein dynamics, at which the theory is not predictive, the Bounce is just a turning point in the past of our Universe. For sufficiently high cut-off energy density the bounce does not significantly alter the thermal history of the Universe as described by the SCM. It can join the current expanding dynamics with a contracting phase. The non-trivial question is the implication that the prebounce phase can have on the expanding branch and, in particular, on the spectrum of metric perturbations.
\\ \indent The present letter has the merit to outline the existence, at least for the isotropic Universe, of a privileged variable linking the LQC approach to the Polymer quantization scheme. In particular, it allows a direct mapping between the Immirzi parameter and the Polymer {\bf lattice size}. 
\\ Furthermore, we show that the horizon paradox can be solved within a bouncing cosmology.
\\ On the other hand, we have shown that the flatness paradox is only weakened, since the issue concerning the fine-tuning of initial conditions (at the Planck time in the SCM $\Omega-1 \sim \mathcal{O}(10^{-60})$) is shifted to the collapsing classical phase of the Universe. However, the explanation of the present value of $\Omega$, which is very close to unity, still requires the specification of a suitable initial conditions at a given time.
\\ \indent Clearly, a precise characterization of a bouncing cosmology requires the study of more general models in the context of PQM, as in \cite{Antonini:2018gdd}. Actually, it is necessary to shed light about the possibility of a semiclassical behavior across the bounce also in the case of a general dynamics, {\it e.g.} having the generality of the Mixmaster model. For instance, in \cite{Crino:2018reo} it is shown that the original Misner idea on the occurrence of physical states with high occupation numbers, even close to the initial singularity, is recovered also in a polymer-like scenario.

\end{document}